\documentclass{amia}
\usepackage{hyperref}
\usepackage{booktabs}
\usepackage{amsfonts}
\usepackage{amssymb}
\usepackage{amsthm}
\usepackage{amsmath}
\usepackage{nicefrac}
\usepackage{multirow}
\usepackage{subfigure}
\usepackage{diagbox}
\usepackage{algpseudocode}
\usepackage{algorithm}
\usepackage[framemethod=tikz]{mdframed}
\usepackage{tcolorbox}
\usepackage{adjustbox}
\usepackage{wrapfig}
\usepackage{bbding}

\begin{document}

\title{Alcohol Intake Differentiates AD and LATE: A Telltale Lifestyle from Two Large-Scale Datasets}

\author{Xinxing Wu, PhD$^a$, Chong Peng, PhD$^b$, Peter T. Nelson, PhD, MD$^a$, Qiang Cheng, PhD$^{a,}${\footnote{Corresponding should be sent to: qiang.cheng@uky.edu.}}}
\institutes{
    $^a$University of Kentucky, Lexington, Kentucky, USA; $^b$Qingdao University, Qingdao, Shandong, China
}

\maketitle

\section*{Abstract}
\vskip -0.1in

\textit{Alzheimer's disease (AD), as a progressive brain disease, affects cognition, memory, and behavior. Similarly, limbic-predominant age-related TDP-43 encephalopathy (LATE) is a recently defined common neurodegenerative disease that mimics the clinical symptoms of AD. At present, the risk factors implicated in LATE and those distinguishing LATE from AD are largely unknown. We leveraged an integrated feature selection-based algorithmic approach, to identify important factors differentiating subjects with LATE and/or AD from Control on significantly imbalanced data. We analyzed two datasets ROSMAP and NACC and discovered that alcohol consumption was a top lifestyle and environmental factor linked with LATE and AD and their associations were differential. In particular, we identified a specific subpopulation consisting of APOE e4 carriers. We found that, for this subpopulation, light-to-moderate alcohol intake was a protective factor against both AD and LATE, but its protective role against AD appeared stronger than LATE. The codes for our algorithms are available at https://github.com/xinxingwu-uk/PFV.}

\section*{Introduction}
\vskip -0.1in

Dementia is a heterogeneous group of disorders with various types and causes~\citep{Kivipelto2018,Causes2020}. Alzheimer’s disease (AD) is its leading form that develops from multifactorial genetic, environmental, and epigenetic causes. More than 35 million people were living with AD and the worldwide societal cost of AD reached \$1 trillion in 2018, with 115.5 million people projected to live with it by 2050~\citep{Mucke2009,Princea2013,Wimo2017}. As an AD-mimic, limbic-predominant age-related TDP-43 encephalopathy (LATE) is a recently-defined type of dementia, with 20\%-50\% of people aged 80 and above having it~\citep{Pete1}. Meanwhile, existing studies have indicated that LATE may appear alone or as a comorbidity with AD~\citep{Pete2,Zhang}. AD usually has a slow onset, and the cognitive deterioration of LATE is even slower than AD; yet, LATE-AD comorbidity  generally causes a more rapid clinical decline than either of them individually~\citep{Boyle}. Their pathological relationship is currently elusive.

Environmental and lifestyle factors play crucial roles in AD and LATE. Managing cardiovascular risk factors, maintaining an active lifestyle (e.g., regular physical, mental, and social activities), and following the Mediterranean diet are associated with a reduced risk for AD or a lower rate of cognitive decline~\citep{Cummings2015}. While the interplay of environmental factors with AD has been investigated~\citep{Mortimer1993,Qiu2010,Imtiaz2014,Sharp2018}, the findings are inconsistent. No effective imaging, fluid, or other types of biomarkers diagnosing LATE or differentiating it from AD is known in a clinical context (in life), nor what lifestyle or environmental factors influence them specifically. Studies on what and how lifestyle and environmental factors are implicated in LATE and AD specifically are urgently needed to help advance our understanding of these types of dementia and provide an accessible target to modify. Considering the size of affected patients and our increasingly aging society, the potential impact of such studies on public health is high.

To find key environmental and lifestyle factors that are associated with AD and LATE specifically, we developed an innovative integrated feature selection-based approach, consisting of Preprocessing, Feature selection, and Validation, abbreviated as PFV. Then, we performed it on two large-scale cohorts, the Religious Orders Study and Memory and Aging Project (ROSMAP) and the National Alzheimer’s Coordinating Center (NACC). Our results revealed that alcohol consumption was a top lifestyle factor linked with LATE and AD and their associations were differential.

In the literature considerable research on the role alcohol consumption plays in cognitive functions and dementia was conducted~\citep{Sabia2018,Rehm2019,Slayday2021}; nonetheless, the findings were mixed or even contradicting. One major limitation of existing studies is that heterogeneity in dementia subtypes was not adequately considered. Moreover, confounders were also insufficiently discerned or controlled. Thus, with the findings by using our PFV, there is an unmet need to study the specific relationship of alcohol consumption with LATE and AD, adjusted for potential confounders.

To address these limitations, we considered AD and LATE specifically and stratified the cohort according to sex, age, and race. We identified a specific subpopulation comprising APOE e4 carriers. For this subpopulation, we found that light-to-moderate alcohol intake was a protective factor against AD and LATE, but its protective role against AD was stronger than LATE.

\section*{Materials and Methods}
\vskip -0.1in

\begin{wrapfigure}{l}{8.5cm}
\vskip -0.2in
\centering
\includegraphics[width=0.53\textwidth]{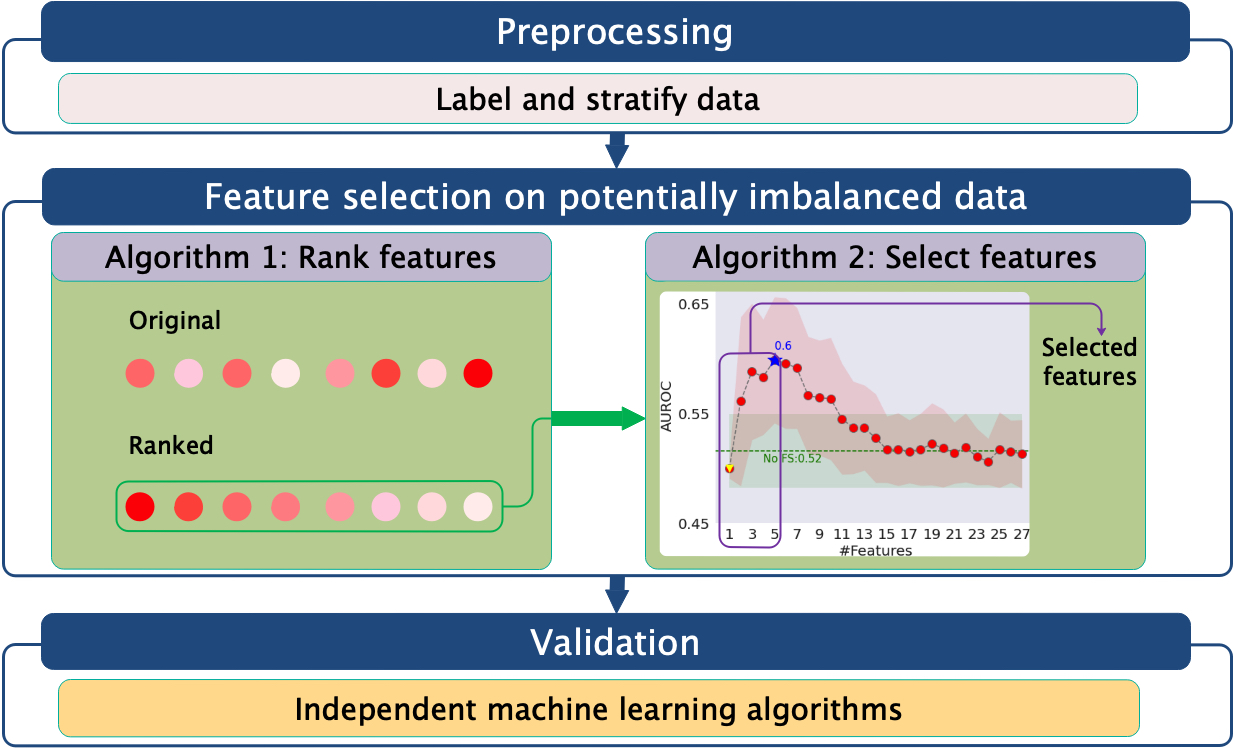}
\vskip -0.1in
\caption{Overall scheme of PFV.}
\vskip -0.1in
\label{architecture}
\end{wrapfigure}

\paragraph{Machine Learning-Based Analysis} The complexity of lifestyle and environmental factors for neurodegeneration and the heterogeneity across subjects are high, which present challenges to the development of early diagnosis tools and effective strategies for preventing dementia~\citep{Myszczynska}. To meet these challenges, this study leveraged machine learning (ML) to identify and rank disease-related environmental and lifestyle factors with large-scale, potentially imbalanced data. Using feature selection (FS) algorithms, we developed an integrated framework, PFV, for ML-based discovery and multi-faceted validation. As summarized in Figure~\ref{architecture}, the steps of PFV include: First, labeling samples and stratifying them based on sex, race, and age; next, performing FS-guided ML analysis to pinpoint a proper group of risk factors from each sub-cohort; finally, verifying the selected risk factors through various independent ML algorithms. More concretely, we explain PFV as follows:

{\sl{Step 1}}. Preprocessing of samples by labeling the examples and deleting rows with missing values. We adopted three existing clinical or neuropathological diagnostic criteria to categorize whether a subject had AD and/or LATE: 1) Braak score~\citep{Braak,Bennett}; 2) CERAD score~\citep{Bennett,Mirra}; 3)TDP-43 stage~\citep{Nag}. We followed the detailed rules in~\citep{Wu} for categorization. Taking ROSMAP as an example, we used the first two scores for annotating subjects with AD and the third measure for LATE. Also, we stratified the cohort data according to sex, race, and age.\\
\begin{wrapfigure}{r}{9.6cm}
\vskip -0.24 in
\centering
\includegraphics[width=0.59\textwidth]{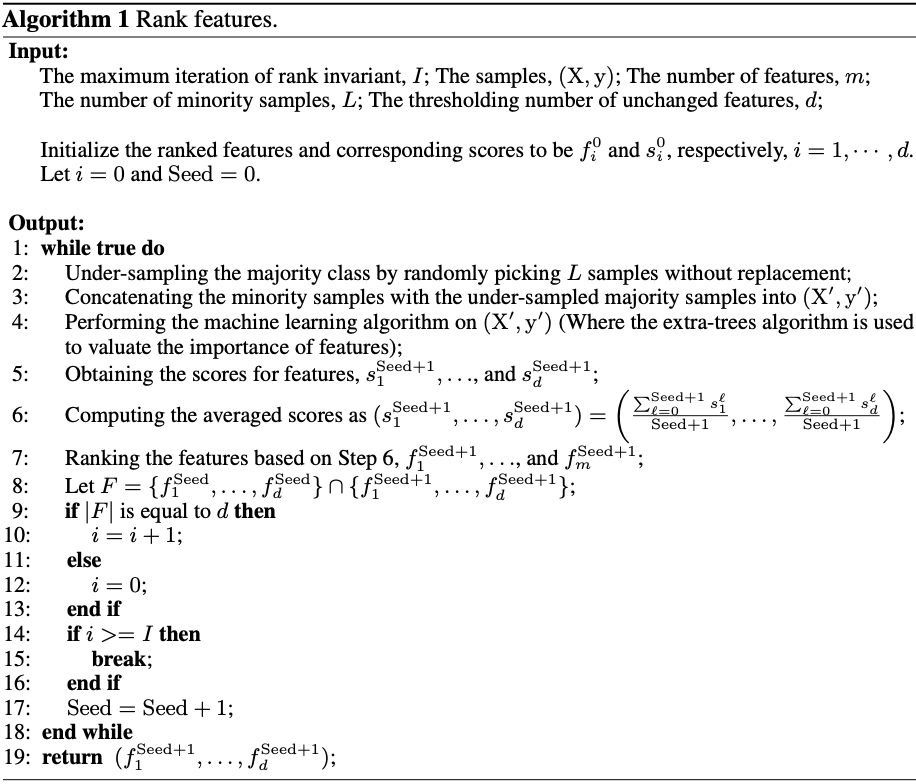}
\label{alg1}
\vskip -0.25 in
\end{wrapfigure}
{\sl {Step 2}}. FS-guided identification of groups of risk factors via Algorithms 1 and 2. This step represents the core discovery phase of our framework. Algorithm 1 was used for ranking all available variables by performing age-, sex-, and race-stratified multivariate risk factor analysis. The ROSMAP data used in our analysis was substantially imbalanced in terms of the sample sizes of controls and cases in many strata. In the literature, ML algorithms were developed to meet the challenges of imbalanced data for classification~\citep{peng2020discriminative, he2009learning} and for FS~\citep{Wu}. However, this ML-based study needed to consider selecting features to account for feature representativeness and inter-correlations and classifying examples jointly. To this end, we developed the two new algorithms. In this paper, the terms of feature, variable, and factor are used exchangeably. If a part of the samples from the majority class(es) had been used, it would have under-utilized the clinically valuable data and led to misleading importance for the variables. On the other hand, as pointed out in~\citep{Wu}, if the majority class(es) had been bootstrapped many times while the remaining minority class(es) had been kept the same, then it would have hardly ensured a sufficient number of used sample subsets for the feature ranking model and, consequently, it would have led to degraded learning performance and made the selected features unstable. To overcome the challenges of imbalanced data, we developed Algorithm 1: First, at each iteration, we generated a new subset by concatenating the minority samples with the under-sampled majority samples, and we applied the FS algorithm to this subset to score the features; next, we averaged the scores obtained from all implemented iterations to get the ranking of all the features. Subsequently, we compared the top $d$ features of the current iteration with the top $d$ features of the last iteration. If the top $d$ features were unchanged for a predefined number of times, we stopped the iteration. Finally, we aggregated the feature importance calculated from all iterations to identify the most informative features in their ability to differentiate distinct classes. In this way, the algorithm can automatically determine the numbers of subsets and iterations to sufficiently analyze and discriminate the imbalanced samples in different classes.

\begin{wrapfigure}{r}{8.0cm}
\vskip -0.2 in
\centering
\includegraphics[width=0.49\textwidth]{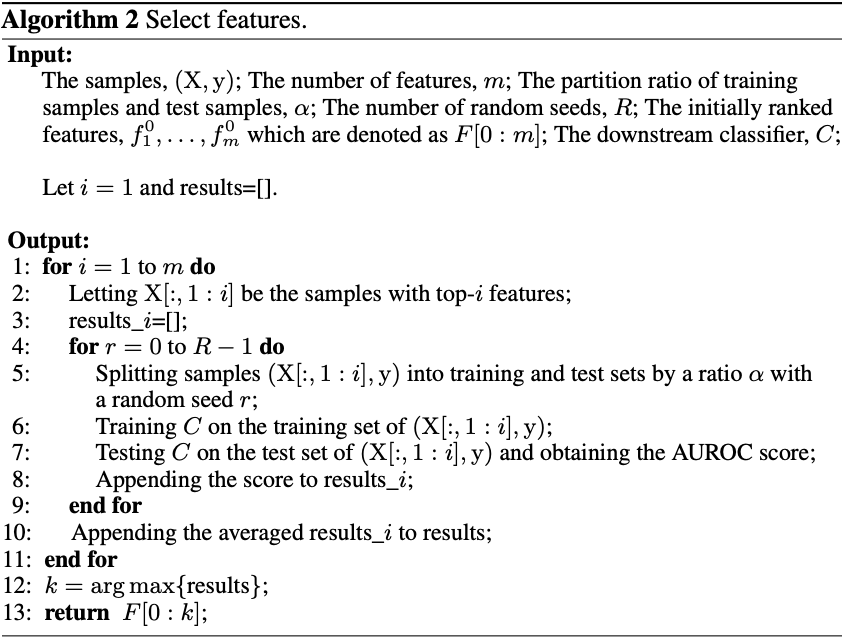}
\label{alg1}
\vskip -0.3 in
\end{wrapfigure}

For selecting important factors from the ranked variables, we further developed Algorithm 2 to account for the representativeness and correlations between variables inspired by~\citep{WuAndCheng1}. First, for the ranked variables by Algorithm 1, we used Algorithm 2 to iteratively construct subsets of risk factors  according to the importance of all variables in descending order{. Then}, we divided the samples with the constructed subset of variables into training and test data, and we trained a classifier based on the training data and tested it to compute the area under the receiver operating characteristic curve (AUROC) score on the test data. Finally, we obtained $m$ AUROC scores, and the one with the maximum AUROC score gave rise to the selected factors. Here, $m$ denotes the number of features.

To guarantee the stability~\citep{WuAndCheng2} of the identified factors, we adopted multiple random seeds for computing AUROC on each subset. Furthermore, we estimated mutual information~\citep{Kraskov2004} for scoring factors in Algorithm 1 and used the extra tree classifier~\citep{Geurts} for selecting factors in Algorithm 2 by the corresponding library functions in Scikit-learn~\citep{Pedregosa2011}.

{\sl {Step 3}}. Validation of selected factors. To assess the efficacy of the identified sets of factors more sufficiently, we adopted further validation by using different independent downstream classifiers. We checked the classification performance with the identified sets of factors using multiple downstream ML algorithms, which were independent of the extra tree classifier in terms of their formulation and property.

In addition, statistical analyses were performed using Python version 3.7.8. Only subjects with complete data were included in the analysis of each specific factor. In ROSMAP, the analysis of variance (ANOVA) test was used to compare alcohol-related variables between groups. The AUROC for accuracy at an optimal cutoff value (i.e., the number of selected features corresponding to the best AUROC) was used to determine factor performance. In  NACC, the chi-square test was used to compare discrete variables, including APOE e4 carrier and alcohol-related variables between groups. Logistic regression analysis tested the associations between alcohol intake-related indexes, APOE e4 carrier, and AD/LATE. $P$-value $<0.05$ was considered to indicate statistical significance.

\section*{Results}
\vskip -0.1in

\paragraph{Dataset Used}{\label{datasetsinfo}}
Data used in this study were ROSMAP and NACC. {\footnote{ROSMAP was obtained via the link https://www.synapse.org/\#!Synapse:syn3219045. NACC was requested via the link https://naccdata.org/}} In ROSMAP, the corresponding clinical indexes and pathological annotations were obtained from the RADC research resource sharing hub. Based on the rules for categorizing~\citep{Wu}, label association was performed on the data for 4 classes: LATE, AD, comorbid LATE and AD (denoted by LATE+AD), and Control. After preprocessing, such as deleting rows with missing values and filtering the features in ROSMAP, we finally obtained 508 samples, each with 33 features including ID and label. Concretely, the preprocessed data contained 47 samples for LATE+AD, 116 samples for pure LATE, 56 samples for pure AD, and 289 samples for Control. In NACC, after filtering alcohol intake-related indexes including ALCDEM and ALCDEMIF, we got 9,256 samples. Specifically, they included 2,117 samples for LATE+AD, 965 samples for pure LATE, 2,242 samples for pure AD, and 3,932 samples for Control.

\begin{figure}[!htbp]
\vskip -0.15in
\centering
\subfigure[All]{
\centering
{
\begin{minipage}[t]{0.2\linewidth}
\centering
\centerline{\includegraphics[height=0.11\textheight]{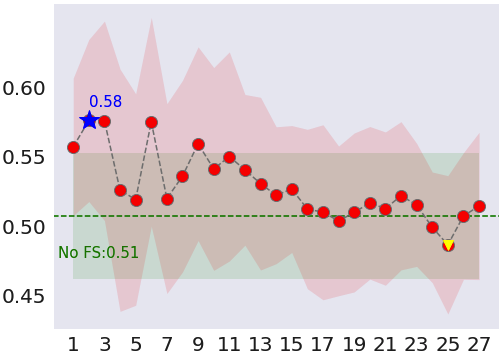}}
\end{minipage}%
}%
}%
\hspace{-0.09in}
\subfigure[Male]{
\centering
{
\begin{minipage}[t]{0.2\linewidth}
\centering
\centerline{\includegraphics[height=0.11\textheight]{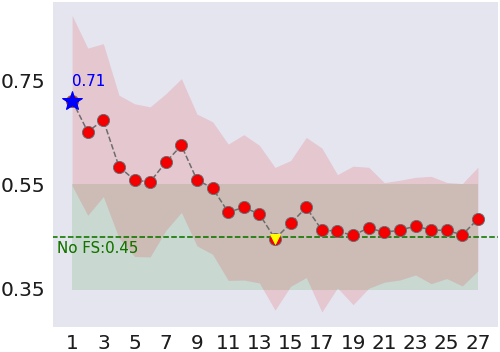}}
\end{minipage}%
}%
}%
\hspace{-0.122in}
\subfigure[Female]{
\centering
{
\begin{minipage}[t]{0.2\linewidth}
\centering
\centerline{\includegraphics[height=0.11\textheight]{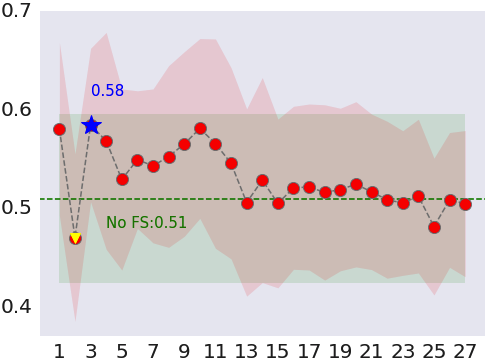}}
\end{minipage}%
}%
}%
\hspace{-0.131in}
\subfigure[White]{
\centering
{
\begin{minipage}[t]{0.2\linewidth}
\centering
\centerline{\includegraphics[height=0.11\textheight]{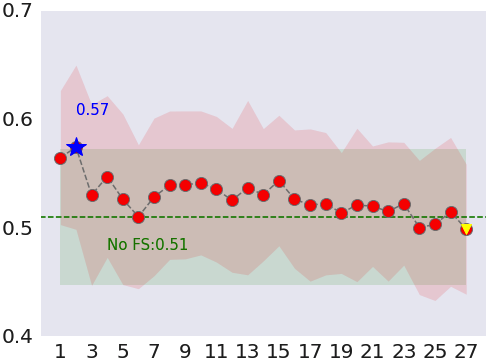}}
\end{minipage}%
}%
}%
\hspace{-0.131in}
\subfigure[$>$85]{
\centering
{
\begin{minipage}[t]{0.2\linewidth}
\centering
\centerline{\includegraphics[height=0.11\textheight]{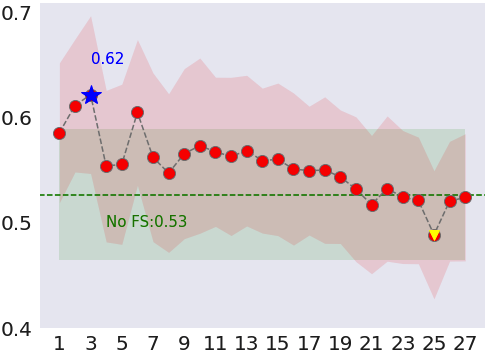}}
\end{minipage}%
}%
}%
\centering
\vskip -0.2in
\caption{AUROC for LATE vs. AD with different numbers of top-ranked features  on different subpopulations (better viewed with color and zoom). The vertical axis is AUROC and the horizontal axis is the number of features.}
\label{res1}
\end{figure}

\begin{table}
\scriptsize
\begin{center}
\vskip 0.15 in
\caption{Several Related Indexes in ROSMAP.}
\vskip -0.1 in
\begin{tabular}{|l|l|l|}
\hline
{\bf Name} & {\bf Symbol}& {\bf Description}\\
\hline
\multirow{2}*{Lifetime daily alcohol intake} & \multirow{2}*{ldai$\_$bl} & The amount of alcoholic drinks consumed per day during the period the participants drank the\\
& & most in their lifetime\\
\hline
Grams of alcohol used per day & alcohol$\_$g$\_$bl & How much alcohol a participant consumed in the past 12 month
\\
\hline
Cognitive resources at 40 & cog$\_$res$\_$age40 & The presence of materials that support cognitive activity in the home at the age of 40\\
\hline
Parental education level&pareduc& Maternal education and paternal education levels\\
\hline
Years of education&educ& The number of years of regular school\\
\hline
Young adult &ya$\_$adult$\_$cogact$\_$freq& Frequency of participation in cognitively stimulating activity as a young adult\\
\hline
\multirow{2}*{Mobility disability} &\multirow{2}*{rosbsum}& Ability to do 3 activities: doing heavy work around the house, walking up and down stairs,\\
& & and walking half a mile without help\\ 
\hline
Lifetime (total) &lifetime$\_$cogact$\_$freq$\_$bl& Frequency of participation in cognitively stimulating activity in lifetime\\
\hline
\multirow{2}*{Instrumental activities of daily living} &\multirow{2}*{iadlsum}& 
Disability using a sum of 8 items adapted from the Duke Older Americans Resources and\\
&& Services project\\
\hline
\multirow{2}*{Basic activities of daily living} &\multirow{2}*{katzsum}& Measuring six basic physical abilities: walking across a small room, bathing, dressing, eating,\\
&&getting from bed to chair, and toileting
\\
\hline
Medical conditions & med$\_$con$\_$sum$\_$bl& A composite measure of 7 medical conditions\\
\hline
\end{tabular}
\label{IndexInROSMAP}
\end{center}
  \vskip -0.4in
\end{table}

\paragraph{Analysis of Risk Factors in ROSMAP}
We performed sex-, race-, and age-stratified ML and statistical analyses in ROSMAP. The details of several related indexes we would use later are listed in Table~\ref{IndexInROSMAP}, and the resulting samples and subpopulation groups are given in Table~\ref{AppendixTable2}. Specifically, Algorithm 1 ranked the features, and Algorithm 2 iteratively performed classification on the ranked features and selected the most informative ones as risk factors. For the parameters in Algorithms 1 and 2, we used $I=500$, $d=27$, $\alpha=0.2$, and $R=50$. We performed the ML analyses in a similar way for different binary classifications and subpopulations, and we obtained the following results:

For LATE vs. AD, we identified 2 risk factors for the whole cohort, 1 risk factor for {the} Male subpopulation, 3 risk factors for {the} Female subpopulation, 2 risk factors for the White subpopulation, and 3 risk factors for {the} $>85$ subpopulation. {U}sing these identified risk factors, the classification accuracy for the whole cohort and the subpopulations of Male, Female, White, and $>85$ improved by about 7\%, 26\%, 7\%, 6\%, and 9\%, respectively, compared to using all the factors.

For LATE vs. Control, we identified 2 risk factors for the whole cohort, 2 risk factors for {the} Male subpopulation, 2 risk factors for {the} Female subpopulation, 7 risk factors for {the} White subpopulation, 6 risk factors for $>85$ subpopulation, and 2 risk factors for $\leqslant 85$ subpopulation. {U}sing these identified risk factors, the classification accuracy for the whole cohort and the subpopulations of Male, Female, White, $>85$, and $\leqslant 85$ improved by about 1\%, 9\%, 4\%, 4\%, 3\%, and 8\%, respectively, compared to using all the factors.

For AD vs. Control, we identified 2 risk factors for the whole cohort, 1 risk factor for {the} Male subpopulation, 2 risk factors for {the} Female subpopulation, 1 risk factor for {the} White subpopulation, and 1 risk factor for {the} $>85$ subpopulation. {U}sing these identified factors, the classification accuracy for the whole cohort and the subpopulations of Male, Female, White, and $>85$ improved by about 5\%, 9\%, 7\%, 5\%, and 9\%, respectively, compared to using all the factors.

For LATE+AD vs. LATE, we identified 3 risk factors for the whole cohort, 5 risk factors for {the} Female subpopulation, 9 risk factors for {the} White subpopulation, and 3 risk factors for {the} $>85$ subpopulation. Using these identified factors improved the classification accuracy for the whole cohort and the subpopulations of Female, White, and $>85$ by about 8\%, 7\%, 3\%, and 8\%, respectively, compared to using all the factors.

\begin{wraptable}{l}{8.3cm}
\scriptsize
\begin{center}
\caption{Preprocessed samples and groups.}
\vspace{-1.55em}
\begin{tabular}{|c|c|c|c|c|}
    \hline
   	{\bf Classes}  & {\bf Subpopulation} & {\bf Class 1}  & {\bf Class 2} &  {\bf Adequacy }\\
    \hline
     \multirow{4}*{\bf LATE (Class 1) } &All &116&56&{\Checkmark}\\
     \cline{2-5}
	\multirow{5}*{\bf vs. }&Male &33&15&{\Checkmark}\\
	\cline{2-5}
	\multirow{6}*{\bf AD (Class 2) }&Female &83&41&{\Checkmark}\\
     \cline{2-5}
     &  Black &7&3&{\XSolidBrush}\\ 
      \cline{2-5}   
     &  White &109&53&{\Checkmark}\\ 
     \cline{2-5}  
     &  $\leqslant 85$ &16&7&{\XSolidBrush}\\ 
      \cline{2-5}   
     &  $>85$ &100&49&{\Checkmark}\\ 
     \hline
     \multirow{6}*{\bf LATE+AD (Class 1)  } &All &47&56&{\Checkmark}\\
     \cline{2-5}
     \multirow{7}*{\bf vs. }& Male& 14&15&{\XSolidBrush}\\
     \cline{2-5}
     \multirow{8}*{\bf AD (Class 2)  }&Female &33&41&{\Checkmark}\\
     \cline{2-5}
     &  Black &2&3&{\XSolidBrush}\\ 
      \cline{2-5}   
     &  White &45&53&{\Checkmark}\\ 
     \cline{2-5}  
     &  $\leqslant 85$ &2&7&{\XSolidBrush}\\ 
      \cline{2-5}   
     &  $>85$ &45&49&{\Checkmark}\\
     \hline
     \multirow{4}*{\bf LATE+AD (Class 1)  } &All &47&116&{\Checkmark}\\
     \cline{2-5}
      \multirow{5}*{\bf vs.  } & Male& 14&33&{\XSolidBrush}\\
     \cline{2-5}
     \multirow{5}*{\bf LATE (Class 2) } &Female &33&83&{\Checkmark}\\
     \cline{2-5}
     &  Black &2&7&{\XSolidBrush}\\ 
      \cline{2-5}   
     &  White &45&109&{\Checkmark}\\  
     \cline{2-5}  
     &  $\leqslant 85$ &2&16&{\XSolidBrush}\\ 
      \cline{2-5}   
     &  $>85$ &45&100&{\Checkmark}\\
     \hline
     \multirow{4}*{\bf LATE+AD (Class 1)  } &All &47&289&{\Checkmark}\\
     \cline{2-5}
      \multirow{5}*{\bf vs. }& Male& 14&117&{\Checkmark}\\
     \cline{2-5}
     \multirow{6}*{\bf Control (Class 2) }&Female &33&172&{\Checkmark}\\
     \cline{2-5}
     &  Black &2&21&{\XSolidBrush}\\ 
      \cline{2-5}   
     &  White &45&267&{\Checkmark}\\ 
     \cline{2-5}  
     &  $\leqslant 85$ &2&88&{\XSolidBrush}\\ 
      \cline{2-5}   
     &  $>85$ &45&201&{\Checkmark}\\  
     \hline
     \multirow{5}*{\bf LATE (Class 1) } &All &116&289&{\Checkmark}\\
     \cline{2-5}
      \multirow{5}*{\bf vs.  }& Male& 33&117&{\Checkmark}\\
     \cline{2-5}
     \multirow{5}*{\bf Control (Class 2)  }&Female &83&172&{\Checkmark}\\
     \cline{2-5}
     &  Black &7&21&{\XSolidBrush}\\ 
      \cline{2-5}   
     &  White &109&267&{\Checkmark}\\
     \cline{2-5}  
     &  $\leqslant 85$ &16&88&{\Checkmark}\\ 
      \cline{2-5}   
     &  $>85$ &100&201&{\Checkmark}\\ 
     \hline
     \multirow{5}*{\bf AD (Class 1)  } &All &56&289&{\Checkmark}\\
     \cline{2-5}
      \multirow{5}*{\bf vs.  }& Male& 15&117&{\Checkmark}\\
     \cline{2-5}
     \multirow{5}*{\bf Control (Class 2) }&Female &41&172&{\Checkmark}\\
     \cline{2-5}
     &  Black &3&21&{\XSolidBrush}\\ 
      \cline{2-5}   
     &  White &53&267&{\Checkmark}\\
     \cline{2-5}  
     &  $\leqslant 85$ &7&88&{\XSolidBrush}\\ 
      \cline{2-5}   
     &  $>85$ &49&201&{\Checkmark}\\ 
    \hline
  \end{tabular}
   \label{AppendixTable2}
\end{center}
  \vskip -0.1in
\end{wraptable}

For LATE+AD vs. AD, we identified 1 risk factor for the whole cohort, 1 risk factor for {the} Female subpopulation, 1 risk factor for {the} White subpopulation, and 3 risk factors for {the} $>85$ subpopulation. Using these identified factors improved the classification accuracy for the whole cohort and the subpopulations of Female, White, and $>85$ by about 12\%, 9\%, 8\%, and 11\%, respectively, compared to using all the factors.

For LATE+AD vs. Control, we identified 5 risk factors for the whole cohort, 3 risk factors for {the} Male subpopulation, 17 risk factors for {the} Female subpopulation, 4 risk factors for {the} White subpopulation, and 2 risk factors for {the} $>85$ subpopulation. Using these identified factors improved the classification accuracy for the whole cohort and the subpopulations of Male, Female, White, and $>85$ by about 8\%, 16\%, 1\%, 9\%, and 6\%, respectively, compared to using all the factors. 

\paragraph{Optimal AUROC and Competitive Classification Performance in ROSMAP}
Taking LATE vs. AD as an example, we show the identified results in Figure~\ref{res1} and Table~\ref{pTable1} (L: LATE, A: AD, and C: Control), respectively. In Figure~\ref{res1}, it is observed that, for most subpopulations, the trained classification model achieved the highest accuracy with the first one or two features. On the other hand, with more features, the classification accuracy typically decreased. Hence, these results indicate that the identified groups of risk factors were significant in that they achieved improved and competitive classification performance when compared with using all factors (i.e., no FS).

\begin{wraptable}{r}{6.8cm}
\scriptsize
\begin{center}
\vskip -0.32 in
\caption{Identified top factors for LATE vs. AD.}
\vspace{-1em}
  \begin{tabular}{|c|l|c|}
    \hline
   	 \multirow{2}*{\bf Subpopulation} & \multirow{2}*{\bf Ranked factors} 			   & {\bf Classification accuracy}\\
	 & &  {\bf (\%, FS vs. No FS)}\\
    \hline
     \multirow{2}*{\bf  Whole  } &1.ldai$\_$bl &\multirow{2}*{58$\uparrow$ (No FS: 51)}  \\ 
     &2.alcohol$\_$g$\_$bl &  \\
     \hline
     {\bf  Male  }&1.ldai$\_$bl &  {71 $\uparrow$ (No FS: 45)} \\
     \hline
     \multirow{3}*{\bf Female} &1.cog$\_$res$\_$age40 &   \multirow{3}*{58 $\uparrow$ (No FS: 51)}\\
      &2.pareduc &    \\
      &3.ldai$\_$bl &   \\  
      \hline
     \multirow{2}*{\bf  White  }&1.ldai$\_$bl &  \multirow{2}*{57 $\uparrow$ (No FS: 51)} \\
     &2.alcohol$\_$g$\_$bl &   \\
     \hline      
     \multirow{3}*{\bf  ${\mathbf{>85}}$  }&1.ldai$\_$bl &  \multirow{3}*{62 $\uparrow$ (No FS: 53)} \\
     &2.alcohol$\_$g$\_$bl &   \\
     &3.educ &  \\ 
     \hline
    {\bf Average}   & \multicolumn{2}{c|}{61.2 $\uparrow$ (No FS: 50.2)}\\
    \hline
  \end{tabular}
\label{pTable1}
\end{center}
\vskip -0.18 in
\end{wraptable}

\begin{figure}
\vskip -0.15in
\centering
\subfigure[LATE+AD vs. LATE]{
\centering
{
\begin{minipage}[t]{0.2\linewidth}
\centering
\centerline{\includegraphics[height=0.108\textheight]{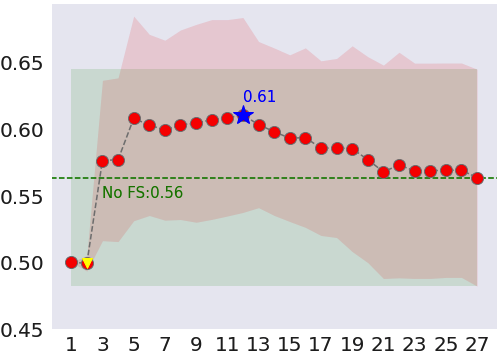}}
\end{minipage}%
}%
}%
\hspace{-0.12in}
\subfigure[LATE+AD vs. AD]{
\centering
{
\begin{minipage}[t]{0.2\linewidth}
\centering
\centerline{\includegraphics[height=0.108\textheight]{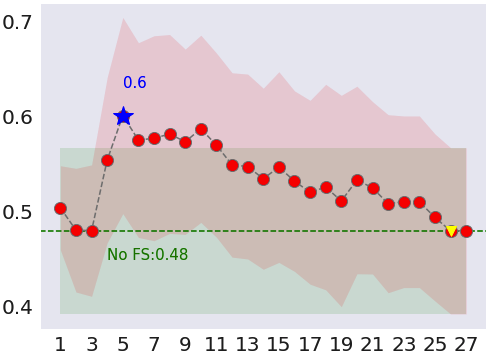}}
\end{minipage}%
}%
}%
\hspace{-0.12in}
\subfigure[LATE vs. AD]{
\centering
{
\begin{minipage}[t]{0.2\linewidth}
\centering
\centerline{\includegraphics[height=0.108\textheight]{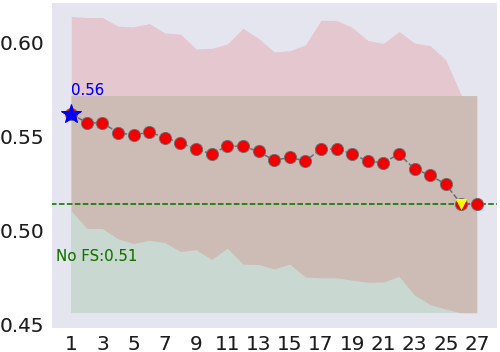}}
\end{minipage}%
}%
}%
\hspace{-0.12in}
\subfigure[LATE+AD vs. LATE]{
\centering
{
\begin{minipage}[t]{0.2\linewidth}
\centering
\centerline{\includegraphics[height=0.108\textheight]{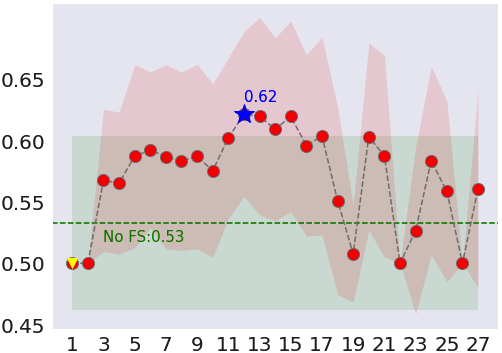}}
\end{minipage}%
}%
}%
\hspace{-0.12in}
\subfigure[LATE+AD vs. AD]{
\centering
{
\begin{minipage}[t]{0.2\linewidth}
\centering
\centerline{\includegraphics[height=0.108\textheight]{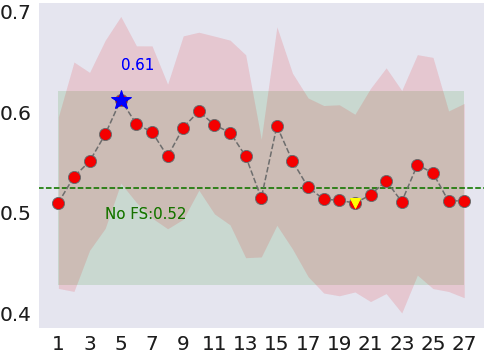}}
\end{minipage}%
}%
}%
\hspace{-0.12in}
\subfigure[LATE vs. AD]{
\centering
{
\begin{minipage}[t]{0.2\linewidth}
\centering
\centerline{\includegraphics[height=0.108\textheight]{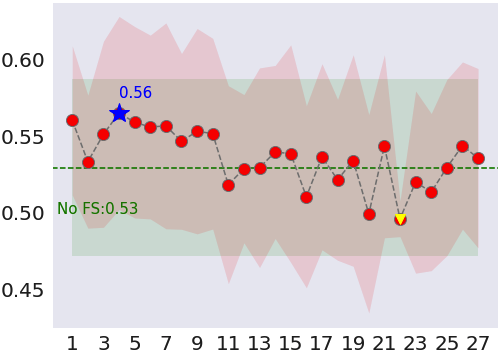}}
\end{minipage}%
}%
}%
\hspace{0.3in}
\subfigure[LATE+AD vs. LATE]{
\centering
{
\begin{minipage}[t]{0.2\linewidth}
\centering
\centerline{\includegraphics[height=0.11\textheight]{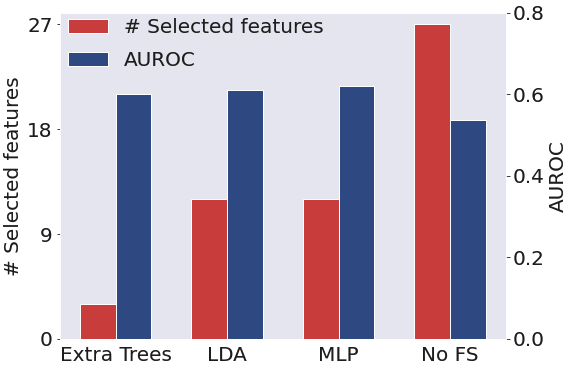}}
\end{minipage}%
}%
}%
\hspace{0.15in}
\subfigure[LATE+AD vs. AD]{
\centering
{
\begin{minipage}[t]{0.2\linewidth}
\centering
\centerline{\includegraphics[height=0.11\textheight]{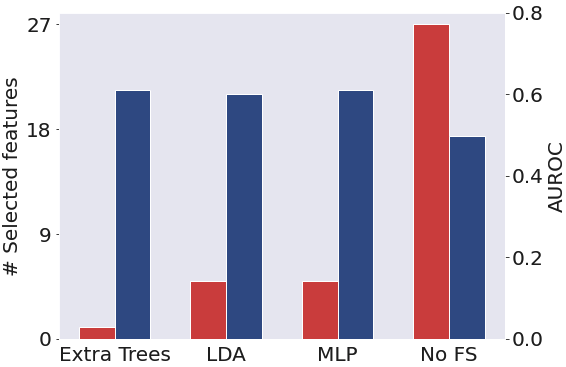}}
\end{minipage}%
}%
}%
\hspace{0.15in}
\subfigure[LATE vs. AD]{
\centering
{
\begin{minipage}[t]{0.2\linewidth}
\centering
\centerline{\includegraphics[height=0.11\textheight]{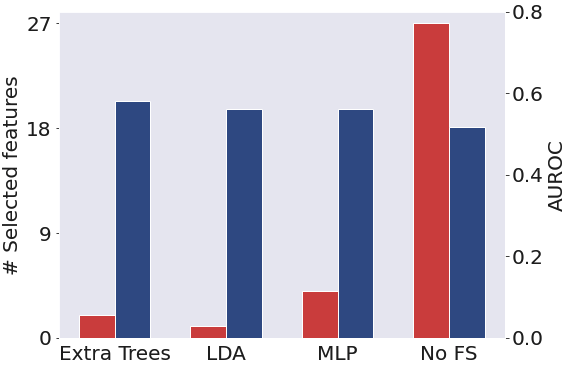}}
\end{minipage}%
}%
}%
\centering
\vskip -0.15in
	\caption{LDA and MLP for the whole cohort. LDA: (a)-(c); MLP: (d)-(f); comparison (in the resulting number of selected features and AUROC) of 3 ML classifiers with No FS: (g)-(i) (better viewed with color and zoom). {In (a)-(f), the vertical axis denotes the AUROC, and the horizontal axis denotes the number of top ranked features.}}
	\label{LDAandMLP}
\end{figure}

\paragraph{Validation with Independent Downstream ML Models in ROSMAP}
To assess the efficacy of identified risk factors, we replaced the extra tree classier in Algorithm 2 with different ML models. We trained and tested the linear discriminant analysis (LDA) and multilayer perceptron (MLP) methods on the ranked features. The obtained results are summarized in Figure~\ref{LDAandMLP}. All these ML models generally achieved the highest accuracy with the top-ranked features, being at least about 7\% more accurate on average than when all factors were used. These improved testing AUROCs on selected factors with independent downstream ML models confirmed the effectiveness of our approach and the identified top factors.

\paragraph{Identification of Risk Factors in ROSMAP}
After analyzing different binary classifications and subpopulations, the identified top factors that could serve as signature patterns were as follows:\\
{\bf LATE vs. AD}: ldai$\_$bl was the common risk factor for the whole cohort and the subpopulations of Male, Female, White, and $>85$; Notably, it was the single most important factor for the Male subpopulation to differentiate LATE vs. AD, among 27 factors (See Table~\ref{pTable1}). Also, alcohol$\_$g$\_$bl was the common risk factor for the whole cohort and the subpopulations of White and $>85$.\\
{\bf LATE vs. Control}: ldai$\_$bl was the common risk factor for the whole cohort and the subpopulations of Male, White, and $>85$; cog$\_$res$\_$age40 was the common risk factor for the whole cohort and the subpopulations of White and $>85$.\\
{\bf LATE+AD vs. LATE}: ldai$\_$bl and ya$\_$adult$\_$cogact$\_$freq were the common risk factors for the whole cohort and the subpopulations of Female, White, and $>85$; rosbsum was the common risk factors for the whole cohort and the subpopulations of White and $>85$.\\
{\bf AD vs. Control}: pareduc was the common risk factor for the whole cohort and the subpopulation of White; lifetime$\_$ cogact$\_$freq$\_$bl was the common risk factor for the whole cohort and the subpopulations of $>85$.\\
{\bf LATE+AD vs{.} Control}: iadlsum was the common risk factor for the whole cohort and the subpopulations of Male, Female, White, and $>85$; katzsum, rosbsum, and ya$\_$adult$\_$cogact$\_$freq were the common risk factors for the whole cohort and the subpopulations of Female and White; katzsum was the common risk factor for the whole cohort and the subpopulations of Female, White, and $>85$; med$\_$con$\_$sum$\_$bl was the common risk factor for the whole cohort and the subpopulations of Male and Female.\\
{\bf For LATE+AD vs. AD}, alcohol$\_$g$\_$bl was the common risk {factor} for the subpopulations of All and $>85$; katzsum was the common factor for the subpopulations of While and $>85$.


\begin{figure}
\centering
\subfigure[All]{
\centering
{
\begin{minipage}[t]{0.2\linewidth}
\centering
\centerline{\includegraphics[height=0.11\textheight]{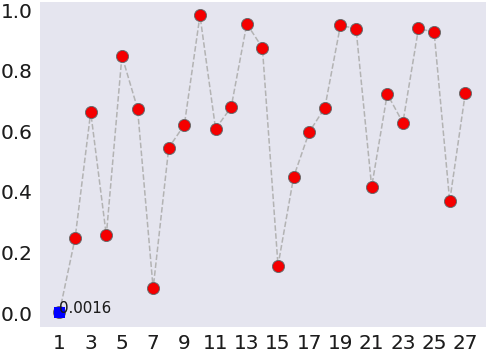}}
\end{minipage}%
}%
}%
\hspace{-0.12in}
\subfigure[Male]{
\centering
{
\begin{minipage}[t]{0.2\linewidth}
\centering
\centerline{\includegraphics[height=0.11\textheight]{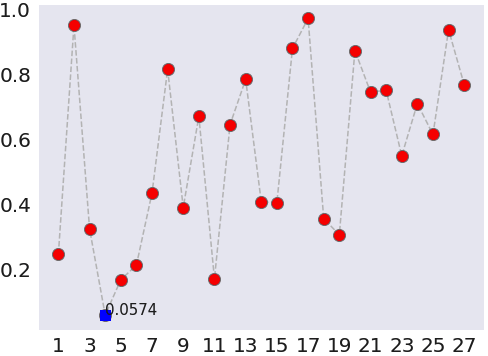}}
\end{minipage}%
}%
}%
\hspace{-0.12in}
\subfigure[Female]{
\centering
{
\begin{minipage}[t]{0.2\linewidth}
\centering
\centerline{\includegraphics[height=0.11\textheight]{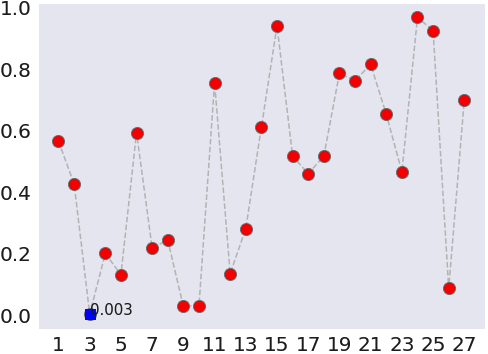}}
\end{minipage}%
}%
}%
\hspace{-0.12in}
\subfigure[White]{
\centering
{
\begin{minipage}[t]{0.2\linewidth}
\centering
\centerline{\includegraphics[height=0.11\textheight]{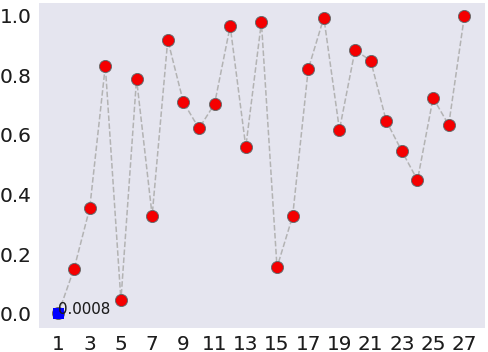}}
\end{minipage}%
}%
}%
\hspace{-0.12in}
\subfigure[$>$85]{
\centering
{
\begin{minipage}[t]{0.2\linewidth}
\centering
\centerline{\includegraphics[height=0.11\textheight]{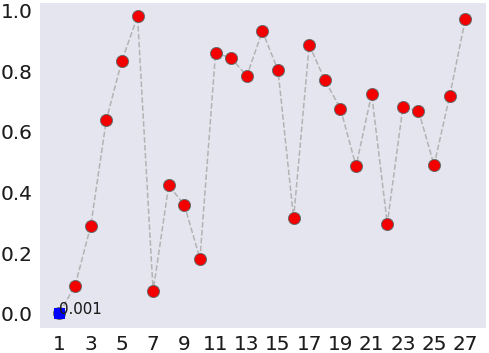}}
\end{minipage}%
}%
}%
\centering
\vskip -0.2in
	\caption{$P$-value of each ranked factor for LATE vs. AD (better viewed with color and zoom). {The vertical axis denotes the $p$-value, and the horizontal axis denotes the ranked feature index.}}
	\label{pv1}
  \vskip -0.1in
\end{figure}

\begin{wraptable}{l}{11.4cm}
\small
\vspace{-1.6em}
\begin{center}
\caption{ANOVA obtained $p$-values for each of {the} two variables on ROSMAP.}
\vspace{-1em}
  \begin{tabular}{|c|c|c|c|c|c|c|}
   \hline
    Variable& L+A vs. L & L+A vs. A &L+A vs. C & L vs. A  & L vs. C  & A vs. C  \\
   \hline
    ldai$\_$bl 			& {\bf $<$0.001} & 0.909 & 0.221 & {\bf 0.002} & {\bf 0.003} & 0.285\\
    \hline
    alcohol$\_$g$\_$bl  & {\bf 0.040}  & 0.563 & 0.433 & 0.245 & 0.119 & 0.991 \\ 
   \hline
  \end{tabular}
   \label{AnovaAlcohol}
\end{center}
\vskip -0.3in
\end{wraptable}

\paragraph{Differential Analysis of Alcohol-related Factors in ROSMAP}
We obtained evidence that the alcohol-related variables had considerably varying, differential importance for the disease types on different strata when adjusting for potential confounders. To confirm the results uncovered by ML algorithms, we perform{ed a} differential analysis with one-way ANOVA for each of the two alcohol intake-related factors, ldai$\_$bl and alcohol$\_$g$\_$bl; see Table~\ref{AnovaAlcohol}. Alcohol intake {was found} statistically significant for the scenarios LATE+AD vs. LATE, LATE vs. AD, and LATE vs. Control. Further, we illustrate $p$-values that were obtained with one-way ANOVA for each of the features ranked by our ML-based algorithm for LATE and AD in Figure~\ref{pv1}. {The top features identified by our ML-based approach had (almost) the smallest $p$-values; for all ranked features, their $p$-values showed an increasing trend.} It is worth noting that our ranking of the factors did not yield a strictly increasing curve in $p$-value, because $p$-values were calculated from linear correlations. In contrast, the ML-based algorithm that we used was the extra tree classifier, capable of capturing the more complex nonlinear relationship. We speculate that ML has {the} potential to more closely model the true relationship between the risk factors and the diseases, which is likely nonlinear due to the high complexity of different forms of dementia.

Overall, the lifestyle factors, such as alcohol intake, daily living activities, and cognitive activities, were top ranked for differentiating four classes: AD, LATE, LATE+AD, and Control. In particular, for alcohol intake, one factor, ldai$\_$bl, emerged as a top differentiating factor for the binary classifications of LATE vs. AD, LATE vs. Control, and LATE+AD vs. LATE, {in} many subpopulations. Besides, another factor, alcohol$\_$g$\_$bl, was also key to differentiating LATE vs. AD and LATE+AD vs. AD.

\begin{table}[!htpb]
\small
\begin{center}
\caption{Alcohol intake-related indexes in NACC.}
  \vskip -0.1in
  \begin{tabular}{|p{0.09\textwidth}|p{0.265\textwidth}|p{0.365\textwidth}|p{0.18\textwidth}|}
   \hline
    {\bf Name} 			 & {\bf Description} & {\bf Numerical value} & {\bf \# Samples}\\
    \hline
    \multirow{3}*{ALCDEM} &  Presumptive etiologic diagnosis of cognitive disorder -- Cognitive impairment due to alcohol abuse
 & No (0, assumed assessed and found not present), Yes (1), and diagnosis of normal cognition (8)  & LATE+AD: 2,117, LATE: 965, AD: 2,242, and Control: 3,932\\
    \hline
    \multirow{3}*{ALCDEMIF} &  Primary, contributing, or noncontributing cause of cognitive impairment -- Alcohol abuse
 & Primary (1), Contributing (2), Non-contributing (3), Cognitively impaired but no diagnosis of impairment due to alcohol abuse (7), and diagnosis of normal cognition (8)  & LATE+AD: 2,117, LATE: 965, AD: 2,242, and Control: 3,932\\
   \hline
  \end{tabular}
   \label{AppendixTable1Add}
\end{center}
  \vskip -0.25in
\end{table}

\begin{wraptable}{l}{12cm}
\small
\begin{center}
  \vskip -0.18in
\caption{$P$-values in NACC for ALCDEM and ALCDEMIF.}
\vspace{-1em}
  \begin{tabular}{|c|c|c|c|c|c|c|}
   \hline
    Variable&  L+A vs. L & L+A vs. A &L+A vs. C & L vs. A  & L vs. C  & A vs. C  \\
    \hline
    ALCDEM				&  {\bf 2.16e-54}  &  0.464  &  {\bf 7.61e-97}   & {\bf 2.86e-52}  & 0.095   &  {\bf 1.87e-97}  \\
    \hline
    ALCDEMIF			& {\bf 7.06e-70}   &  0.297 &  {\bf 5.06e-130}   &  {\bf 6.72e-70} &  {\bf 0.037}  &   {\bf 2.77e-134} \\
   \hline
  \end{tabular}
   \label{nacc1}
\end{center}
  \vskip -0.3in
\end{wraptable}

\paragraph{Confirmation of Findings in NACC} 
We considered two alcohol-related indexes, ALCDEM and ALCDEMIF, which were related to the diagnosis of cognitive impairment due to alcohol abuse in NACC; see Table~\ref{AppendixTable1Add}. We performed cross-cohort independent validation of our findings for the two indexes by using the chi-square test. The results are summarized in Table~\ref{nacc1}. It is observed that the two variables were statistically significant for 4 binary classification scenarios: LATE+AD vs. LATE, LATE+AD vs. Control, LATE vs. AD, and AD vs. Control. However, they could not differentiate LATE+AD vs. AD; and ALCDEM could not work well for LATE vs. Control. The APOE e4 genetic variation was related to LATE and AD~\citep{Slayday2021,adam2021}, so we further examined the relationships of LATE and AD with ALCDEM/ALCDEMIF and APOE e4 carrier. First, by using the chi-square test, we analyze{d} the relationship between alcohol and APOE e4 carrier status (see Table~\ref{nacc11}). It is observed that there {were} statistically significant ($p$-values$<$0.05) associations between ALCDEM/ALCDEMIF and APOE e4 in all 12 scenarios except for ALCDEMIF in the scenario of LATE vs. Control.

\begin{wraptable}{l}{12cm}
\small
\begin{center}
  \vskip -0.18in
\caption{$P$-values in NACC for APOE e4 carrier with ALCDEM /ALCDEMIF.}
\vspace{-1em}
  \begin{tabular}{|c|c|c|c|c|c|c|}
    \hline
    Variable&  L+A vs. L & L+A vs. A &L+A vs. C & L vs. A  & L vs. C  & A vs. C  \\
    \hline
    ALCDEM				&  {\bf 7.17e-14}  &  {\bf 7.06e-12}  &  {\bf 9.23e-35}   & {\bf 8.39e-10}   &  {\bf 0.042}  &  {\bf 8.59e-31}  \\
    \hline
    ALCDEMIF				&  {\bf 2.00e-17}   &  {\bf 5.40e-15}   &   {\bf 6.84e-41}   &  {\bf 9.23e-14}   & 0.083   &   {\bf 8.37e-38}  \\
   \hline
  \end{tabular}
   \label{nacc11}
\end{center}
  \vskip -0.2in
\end{wraptable}

Then, we studied the relations of alcohol intake and APOE e4 with AD and/or LATE. We used the logistic regression for each alcohol variable, adjusted for APOE e4 carrier status in Tables~\ref{nacc2} and~\ref{nacc7} (CI: confidence interval). As is well known, APOE e4 is a potent risk factor for AD, which was re-confirmed by our results. For instance, for AD vs. Control, with ALCDEM and ALCDEMIF as covariates, the odds ratios (OR) for APOE e4 carriers were 2.010 and 5.068, respectively, indicating that APOE e4 carriers were about 2 times and 5 times likely to be affected by AD compared to Control. Also, it is found that APOE e4 was more of a risk factor for AD than for LATE. For instance, for AD vs. LATE, with ALCDEM and ALCDEMIF as independent covariates, the OR were 4.998 and 3.556, respectively, indicating that APOE e4 carriers had a risk for AD about 5 times and 3.6 times that for LATE. In the scenario of LATE vs. LATE+AD, with ALCDEM and ALCDEMIF as independent covariates, the OR were 0.205 and 0.250, respectively, indicating that APOE e4 carriers were about 3.9 times and 3 times more susceptible to LATE+AD than to pure LATE. 

 \begin{table}[!htbp]
 \vskip -0.1in
  \centering
  \footnotesize
  \caption{ALCDEM+APOE e4 carrier vs. LATE and/or {AD} in NACC by logistic regression analysis.}
  \vskip -0.1in
  \begin{tabular}{|c|c|c|c|c|c|c|c|c|c|c|}
   \hline
   	{\multirow{3}*{\diagbox{Response}{$p$-value}{Variable}}} & \multicolumn{9}{c|}{APOE (e4 carrier) and ALCDEM}\\
   \cline{2-10}
     & {\multirow{2}*{LLR $p$-value}} & \multicolumn{4}{c|}{APOE}   & \multicolumn{4}{c|}{ALCDEM} \\
	\cline{3-10}
     &  &  OR &  $p$-value  & Coef. & [97.5\% CI] & OR & $p$-value  & Coef. & [97.5\% CI] \\
   \hline
 	L vs. L+A 									&  {\bf 1.91e-90}   &  0.205  & {\bf $<$0.001} & -1.587&[-1.720,-1.454]  &   1.177  &  {\bf $<$0.001}  &0.163&  [0.135,0.191]\\
    \hline
    A vs. L+A  									&  1.000   &  \slash  &  \slash  &\slash&\slash&  \slash   &   \slash &  \slash   &   \slash \\
    \hline
   	C vs. L+A  									&  {\bf 3.57e-137}   &  0.5154  & {\bf $<$0.001}&-0.6643&[-0.758,-0.570]   &  1.327   &  {\bf $<$0.001}  &0.283&[ 0.261,0.305]\\
    \hline
    A vs. L  									&  {\bf 2.05e-73}   &  4.998  &  {\bf $<$0.001}  &1.609&[1.477,1.741]&   0.865  &  {\bf $<$0.001}  & -0.146& [-0.172,-0.119]\\
    \hline
    L vs. C  									&  1.000   &  \slash  & \slash   &   \slash & \slash  &   \slash  &   \slash  &  \slash   &   \slash\\
    \hline
    A vs. C  									&  {\bf 1.01e-149}   & 2.010   & {\bf $<$0.001}   &0.698&[0.605,0.791]&   0.766  &  {\bf $<$0.001} &-0.266&  [-0.287,-0.246]\\
   \hline
  \end{tabular}
  \label{nacc2}
 \end{table}

\begin{table}[!htbp]
  \centering
  \footnotesize
  \caption{ALCDEMIF+APOE e4 carrier vs. LATE and/or AD in NACC by logistic regression analysis.}
  \vskip -0.1in
  \begin{tabular}{|c|c|c|c|c|c|c|c|c|c|c|}
   \hline
   	{\multirow{3}*{\diagbox{Response}{$p$-value}{Variable}}} & \multicolumn{9}{c|}{APOE (e4 carrier) and ALCDEMIF}\\
   \cline{2-10}
     & {\multirow{2}*{LLR $p$-value}} & \multicolumn{4}{c|}{APOE}   & \multicolumn{4}{c|}{ALCDEMIF} \\
	\cline{3-10}
     &  &  OR &  $p$-value  & Coef. & [97.5\% CI] & OR & $p$-value  & Coef. & [97.5\% CI]\\
   \hline
 	L vs. L+A 									&  {\bf 6.41e-59}   & 0.250   & {\bf $<$0.001}    &-1.386&[ -1.549,-1.222]&    0.997  &  0.675   &-0.003&[-0.018,0.012]\\
    \hline
    A vs. L+A  									&  0.081   &  0.892  &  0.077  &-0.115&[-0.242,0.012]&  1.021   &   {\bf 0.004}  &0.021&[0.007,0.035]\\
    \hline
    C vs. L+A  									&  {\bf 8.28e-206}   &  0.175  &  {\bf $<$0.001}  & -1.741&[-1.861,-1.621]& 1.214   &  {\bf $<$0.001}  &0.194&[0.182,0.206] \\
    \hline
    A vs. L  									&  {\bf 4.63e-49}   &  3.556  & {\bf $<$0.001}   &1.269&[1.107,1.430]&   1.025  &   {\bf 0.001} &0.024&[0.010,0.039]\\
    \hline
    L vs. C  									&  {\bf 9.61e-08}   &  1.450  &  {\bf $<$0.001}  &0.372&[0.216, 0.528]&   0.823  &   {\bf $<$0.001}  & -0.195&[-0.207,-0.183]\\
    \hline
    A vs. C  									&  {\bf 3.44e-189}   &  5.068  &  {\bf $<$0.001}  &1.623&[1.507,1.739]&  0.841   &   {\bf $<$0.001}  & -0.173&[-0.184,-0.162]\\
   \hline
  \end{tabular}
  \label{nacc7}
  \vskip -0.2in
 \end{table}
 
For the subjects in Control class in Tables~\ref{nacc2} and~\ref{nacc7}, we first calculated the average value of ALCFREQ to be about 2.34. {\footnote{{ALCFREQ denotes that, during the past three months, how often the subject had at least one drink of any alcoholic beverage: 0 is for less than once a month, 1 for about once a month, 2 for about once a week, 3 for few times a week, and 4 for daily or almost daily.}}} {Thus, we could} observe significant protective roles of light-to-moderate alcohol intake against AD, LATE, or LATE+AD, adjusted for APOE e4. ALCDEM and ALCDEMIF show{ed} significant ($p$-value$<$0.001) OR=0.766 and OR=0.841 in comparing AD vs. Control, and thus {were} associated with about 23\% and 16\% decrease in risk for AD, respectively, adjusted for APOE e4 carrier status. ALCDEM and ALCDEMIF show{ed} significant ($p$-value$<$0.001) OR=1.327 and OR=1.214 in comparing Control vs. LATE+AD, and thus {were} associated with about 25\% and 18\% decrease in risk for LATE+AD, respectively, adjusted for APOE e4 carrier status. ALCDEMIF show{ed} a significant ($p$-value$<$0.001) OR=0.823 in comparing LATE vs. Control, and thus {were} associated with about 18\% decrease in risk for LATE, adjusted for APOE e4 carrier status. ALCDEM show{ed} a significant ($p$-value$<$0.001) OR=0.865 in comparing AD vs. LATE, and thus {were} associated with about 14\% lower risk for AD than for LATE, adjusted for APOE e4 carrier status. Essentially, in a cross-cohort independent verification fashion, our findings in ROSMAP were confirmed in NACC. For other alcohol-related variables in NACC, we also confirmed the findings. Due to space limit, we will report the results for other alcohol-related variables in a separate paper.

\section*{Discussion}
\vskip -0.1in

Existing studies~\citep{Mortimer1993,Sabia2018,Rehm2019,Andrew,Ko2018,AlcoholandDementia,Marshall2013,Marshall2020,Katzman1993,Mezencev2020,Tan2008} have shown that {several lifestyle} factors, including alcohol, daily living activities, and cognitive activities, are closely linked to AD. In particular, research on the role of alcohol consumption in cognitive functions and dementia was conducted~\citep{Sabia2018,Rehm2019,Slayday2021}; nonetheless, their findings were mixed or even contradicting. 

For the first time, our results suggest that alcohol consumption is associated with LATE and its comorbidity with AD. The existing work closely related to our study is~\citep{Slayday2021}, which tested the interaction of alcohol with APOE e4 on 7 cognitive measures for a middle-aged cohort without considering dementia types such as AD and LATE. The conclusion of~\citep{Slayday2021} was that the e4 allele might increase vulnerability to the deleterious effects of heavy alcohol intake, while beneficial effects of light or moderate alcohol intake were not seen. Because most LATE and AD cases are typically for the old-aged, the conclusion of~\citep{Slayday2021} cannot be generalized to AD and LATE. In contrast, we examine{d} the old-aged cohorts to dissect the relationship of alcohol consumption with dementia types AD and LATE specifically, thus obtaining starkly different findings from~\citep{Slayday2021}. Another related work~\citep{Yan2021} studied the old-aged male cohort for cognitive functions and clinical dementia diagnosis, without considering APOE or pathological or etiological distinction between AD and LATE. In addition, recent work~\citep{adam2021} tested 4 genetic variants for their links with LATE and AD adjusted for APOE e4, with no environmental or lifestyle factors considered.

To our knowledge, we are the first to study the association of alcohol with LATE/AD specifically, and our findings indicate that light-to-moderate alcohol consumption played differential protective roles against AD, LATE, and AD+LATE for APOE e4 carriers in ROSMAP and NACC.

\vskip -0.15in
\section*{Conclusion{s}}
\vskip -0.1in

This study explored environmental and lifestyle factors, including lifestyle, education, medical history, diets, and cognitive activity to differentiate between LATE, AD, and Control. We introduced an ML-based approach for identifying important environmental and lifestyle factors in LATE and/or AD. Using the proposed approach, we found that alcohol intake was linked with AD and LATE, which was first identified on a discovery dataset ROSMAP and then further validated on an independent dataset NACC. Notably, for the first time, our results suggest that alcohol consumption was associated with LATE and AD often as a top lifestyle and environmental factor, but its associations with LATE and AD were differential. In particular, we identified a specific subpopulation comprising APOE e4 carriers. We discovered that, for this particular subpopulation, light-to-moderate alcohol intake was a protective factor against AD and LATE, but its protective role against AD was greater than LATE. Our study will potentially contribute to understanding the different lifestyle and environmental factors implicated in LATE and/or AD. 

\subparagraph{Acknowledgments}
\vskip -0.1in

This work was partially supported by the NIH grants R21AG070909, R56NS117587, R01HD101508, and ARO W911NF-17-1-0040. And the results published here are in whole or in part based on data obtained from the AD Knowledge Portal and NACC. 
\vskip -0.18in

\makeatletter
\renewcommand{\@biblabel}[1]{\hfill #1.}
\makeatother

\vskip -0.2in
\begin{center}
\bibliographystyle{vancouver}
\bibliography{references}  
\end{center}

\end{document}